\documentclass[5p,12pt]{elsarticle}

\usepackage{graphicx}
\graphicspath{{FiguresWeb/}}

\usepackage{algorithm}
\usepackage{algorithmic}
\usepackage{amsmath}
\usepackage{amssymb}
\usepackage{caption}
\usepackage{mathtools}
\usepackage{fixltx2e}
\usepackage{subfig}
\usepackage{url}
\usepackage{etoolbox}
\usepackage{hyperref}
\usepackage[all]{hypcap}
\usepackage{nomencl}

\AtBeginEnvironment{algorithmic}{\footnotesize} 
\makenomenclature

\biboptions{comma,square}

\journal{J. Parallel Distrib. Comput.}

\begin{document}

\begin{frontmatter}

\title{Solving the Resource Constrained Project Scheduling Problem Using the Parallel Tabu Search Designed for the CUDA Platform\footnotemark}

\author{Libor Bukata\corref{cor}}
\ead{libor.bukata@cvut.cz}
\author{P\v{r}emysl \v{S}\r ucha}
\ead{premysl.sucha@cvut.cz}
\author{Zden\v{e}k Hanz\'{a}lek}
\ead{zdenek.hanzalek@cvut.cz}

\cortext[cor]{The corresponding author.}
\address{Department of Control Engineering, the Czech Technical University in Prague,\\Karlovo n\'{a}m\v{e}st\'{i}~13, 121 35 Prague 2, the Czech Republic}

\begin{abstract}
In the paper, a parallel Tabu Search algorithm for the Resource Constrained Project Scheduling Problem is proposed.
To deal with this NP-hard combinatorial problem many optimizations have been performed.
For example, a resource evaluation algorithm is selected by a heuristic and an effective Tabu List was designed.
In addition to that, a capacity-indexed resource evaluation algorithm was proposed 
and the GPU (Graphics Processing Unit) version uses a homogeneous model to reduce the required communication bandwidth.
According to the experiments, the GPU version outperforms the optimized parallel CPU version with respect to the computational time and the quality of solutions.
In comparison with other existing heuristics, the proposed solution often gives better quality solutions.

\vspace{1em}
\noindent
\emph{Cite as}: Libor Bukata, Premysl Sucha, Zdenek Hanzalek, Solving the Resource Constrained
Project Scheduling Problem using the parallel Tabu Search designed for the CUDA platform,
\emph{Journal of Parallel and Distributed Computing}, Volume 77, March 2015,
Pages 58-68, ISSN 0743-7315, \url{http://dx.doi.org/10.1016/j.jpdc.2014.11.005}.

\vspace{1em}
\noindent
\emph{Source code}: 
\url{https://github.com/CTU-IIG/RCPSPCpu}, \url{https://github.com/CTU-IIG/RCPSPGpu}
\end{abstract}

\begin{keyword}
Resource Constrained Project Scheduling Problem \sep Parallel Tabu Search \sep CUDA \sep homogeneous model \sep GPU
\end{keyword}

\end{frontmatter}

\footnotetext{\copyright 2017. This manuscript version is made available under the CC-BY-NC-ND 4.0 license \url{http://creativecommons.org/licenses/by-nc-nd/4.0/}.}

\section{Introduction}
\label{lab:intro}

The \emph{Resource Constrained Project Scheduling Problem} (RCPSP),
which has a wide range of applications in logistics, manufacturing and project management \cite{Fu2012}, 
is a universal and well-known problem in the operations research domain.
The problem can be briefly described using a set of activities and a set of precedence constraints describing the relationships among activities.
Each activity requires a defined amount of the resources and every resource has a limited capacity.
The objective is to find the best feasible schedule according to a criterion.
The RCPSP was proved to be NP-hard in the strong sense when the criterion is makespan \cite{Blazewicz198311}.
For that reason only small instances (approximately up to 30 activities) can be reliably solved by exact methods like Branch \& Bound \cite{Chaleshtarti6115033},
therefore a heuristic or a meta-heuristic is required to solve the problem satisfactorily.

In recent times, there is an increased interest in using graphics cards to solve difficult combinatorial problems (e.g. \cite{Delevacq201352,Czapinski2012,Bozejko20122020}),
since a modern graphics card is usually much more powerful than a current multi-core CPU.
Although the graphics cards have some restrictions (e.g. a high-latency global memory access),
the new GPU architectures like Kepler and Fermi can significantly reduce these bottlenecks.
As a consequence modern GPUs are applicable to the problems which were solvable only on CPUs previously.

Not only the high computational power makes graphics cards attractive to researchers and practitioners,
but also the mature Nvidia CUDA framework which enables us to create GPU programs in an effective and relatively easy way
since it extends standard languages like C/C++ by adding GPU specific functions and language keywords.
Nevertheless, the CUDA is only designed for the Nvidia graphics cards.

From the implementation point of view, there are two models.
The first one is called a \emph{homogeneous model} where all required data structures are stored in a GPU at the beginning of an algorithm and the results are read at the end of the algorithm.
There is no communication between the CPU and the GPU during the computations.
The second approach is a \emph{heterogeneous model}.
The main logic of an algorithm runs on the CPU and the GPU is used only for the most computationally intensive tasks.
The disadvantage of the heterogeneous model is the frequent communication during computations, therefore, the communication bandwidth can state a bottleneck.
However, the heterogeneous model is usually simpler to implement.

\subsection{Related works}

The Tabu Search meta-heuristic was proposed by Glover in 1986~\cite{Glover1986}.
Hundreds of publications have been written since that time.
The basic concept of the TS meta-heuristic is clarified in Gendreau \cite{TS_INTRO}.
The author has described the basic terms of the TS, as a Tabu List (TL), aspiration criteria, diversification, intensification, etc.

From the Tabu Search parallelization point of view James et al. \cite{James2009810} proposed a sophisticated solution.
The authors use a circular buffer where the size of the buffer is equivalent to the number of the started threads (often the number of CPUs cores).
Every location (i.e. an index of a thread) has different parameters (a tabu tenure, stopping criteria).
At the beginning of the search each thread initializes its own location by a short TS operator, i.e. a modified version of the Taillard's robust tabu search.
Then the asynchronous parallel tabu search is started.
Every thread independently reads a solution and parameters from the location, possibly makes a diversification, runs the TS operator on the solution,
and writes back and sets an UPDATE flag if an improving solution is found. After that the thread location index is circularly incremented.
Diversification takes place if the read solution does not have the UPDATE flag set.
Every best global solution is copied to half of the locations of the circular buffer to propagate elite solutions.
Since the circular buffer is shared by many threads, the access has to be as short as possible and the locations have to be protected by critical sections.

Relatively many authors try to use a GPU for solving combinatorial problems.
For example, Czapi\'{n}ski and Barnes \cite{FlowShopGPU} implemented a GPU version of TS to solve the Flowshop Scheduling Problem (FSP).
The success of the implementation illustrates an achieved speedup against the CPU version.
The GPU version was up to 89.01 times faster than the CPU (Intel Xeon 3.0 GHz, 2 GB memory, Nvidia Tesla C1060 GPU).
Nevertheless, the quality of solutions was not investigated.

The Flowshop Scheduling Problem was also solved by Zaj\'{i}\v{c}ek and \v{S}\r ucha \cite{SuchaZajicek2012}.
The authors implemented a~GPU version of an island based genetic algorithm.
Islands are used for migration of individuals among sub-populations,
where each sub-population is a subset of solutions and an individual corresponds to a specific solution.
Sub-population can be evaluated, mutated and crossed over independently of other sub-populations,
therefore, huge parallelization can be achieved.
It should be noted that a homogeneous model was used.
The maximal speedup against the CPU was 110 for 100 activities and 5 machines (AMD Phenom II X4 945 3.0 GHz, Nvidia Tesla C1060).

Czapi\'{n}ski \cite{Czapinski2012} proposed a Parallel Multi-start Tabu Search for the Quadratic Assignment Problem.
The main idea is to start several Parallel Tabu Search instances with different parameters and initial solutions.
All Tabu Search instances should terminate approximately at the same time since the synchronization is required to get the most promising solutions.
When a stop criterion is met, the modified solutions are read back and the most promising solutions are used as the initial solutions in the next run.
The Tabu Search instance runs entirely on the GPU, therefore communication overheads are reduced to minimum.
The achieved results reveal the effectiveness of the implementation since Nvidia GTX 480 is up to 70 times faster than a six-core Intel Core i7-980x 3.33 GHz.


Hofmann et al. \cite{Hofmann2013} investigated the suitability of graphics cards for genetic algorithms.
The authors selected two problems to solve, namely the Weierstrass function minimization and the Traveling Salesman Problem.
 The first problem can be very effectively implemented on the GPU since the Weierstrass function 
is comprised of floating-point operations and trigonometric functions that are directly supported by the GPU hardware.
In contrast to the first problem the second problem was not tailor-made for the GPU, therefore, the multi-core CPU was able to compete with the GPU.
The authors suggest that all parts of a genetic algorithm should be performed on the Fermi or newer GPUs (i.e. homogeneous model).

Boyer et al. \cite{Boyer201242} used dynamic programming to solve the knapsack problem on a GPU.
An effective data compression was proposed to reduce memory occupancy.
The achieved results show that the Nvidia GTX 260 graphics card was up to 26 faster than Intel Xeon 3.0 GHz.


The above mentioned combinatorial problems have something in common.
The solution evaluation is quite simple since it is usually a ``simple sum''.
On the other hand, the RCPSP requires much more complicated schedule evaluation methods and data structures.

\subsection{Contribution and Paper Outline}

The proposed solution is the first known GPU algorithm for the RCPSP.
The performed experiments revealed that the GPU outperforms the CPU version in both performance speedup and the quality of solutions.
This is possible thanks to an effective schedule evaluation and a GPU-optimized Simple Tabu List.
In addition, the required data transfers are reduced to minimum due to the homogeneous model.
Our Parallel Tabu Search is able to outperform other Tabu Search implementations in the quality of the resulting solutions.

The paper is structured as follows:
The following section introduces the RCPSP mathematical formulation and notation.
The Tabu Search meta-heuristic is briefly described in Section~\ref{sec:briefTSDesc}.
Our proposed Parallel Tabu Search algorithm for the CUDA platform is described in detail in Sections~\ref{sec:PTSG}, \ref{sec:scheduleEvaluation}, and \ref{sec:PTS_CUDA}.
The performed experiments are located in Section \ref{sec:ExRes} and the last section concludes the work.

\section{Problem Statement}\label{sec:problemStatement}

According to the standard notation, classification for the RCPSP is $\textrm{PS}|prec|C_{max}$ \cite{Brucker19993}.
A project can be described as follows:
There is a set of activities $V = \{0, \dots, N-1\}$ with durations $D = \{d_0, \dots, d_{N-1}\}$ where \emph{N} is the number of activities.
There are two dummy activities $0$ and $N-1$ such that $d_0 = d_{N-1} = 0$.
Activity $0$ is a predecessor of all other activities and activity $N-1$ is the end activity of a project.
A schedule of the RCPSP can be represented as a vector of activities' start times $S = \{s_0, \dots, s_{N-1}\}$ where $s_i \in \mathbb{N}$.
Alternatively, a schedule can be expressed as an order of activities $W = \{w_0, \dots, w_{N-1}\} \in \mathcal{W}$ where $w_u$ is the $u$-th activity of the schedule and $\mathcal{W}$ is a set of all feasible solutions.

The RCPSP can be represented as Direct Acyclic Graph $G(V,E)$ where nodes $V$ are activities and edges $E$ are precedence relations.
If there is edge $(i,j) \in E$ then $s_i+d_i \leq s_j$ since activity $j$ has to be scheduled after activity $i$.

Each activity requires some amount of renewable resources.
The number of project resources is denoted as \emph{M} and a set of resources capacities is $\mathcal{R} = \{R_0, \dots, R_{M-1}\}$ where $R_k \in \mathbb{N}$.
Maximal resource capacity $R_{max}$ is equal to $\max_{k=0}^{M-1} R_k$.
Activity resource requirement $r_{i,k} \in \mathbb{N}$ means that activity $i$ requires $r_{i,k} \leq R_k$ resource units of resource $k$ during its execution.
As $s_i$ and $d_i$ values are positive integers, the resulting schedule length $C_{max}$ (i.e. the project makespan) will also be an integer as well.

A lower bound of the project makespan can be found by neglecting resources.
For each activity $i \in V$ all outgoing edges $(i,j) \in E$ are weighted by its duration $d_i$.
The longest path from $0$ to $N-1$ in graph $G(V,E)$ corresponds to the \emph{critical path}.
Its length is equal to the optimal project makespan on the condition that all resources have an unlimited capacity.

\subsection{Mathematical Formulation}

\noindent
\begin{align}
	\text{minimize} & \;C_{max} \label{eq:f1} \\
	\text{s.t.}\quad & \;C_{max} = \displaystyle \max_{\forall i \in V} \left(s_i+d_i\right) \label{eq:f2} = s_{N-1} \\
	& \;s_j \geq s_i + d_i \qquad \forall (i,j) \in E \label{eq:f3} \\
	& \;\displaystyle\max_{t=0}^{C_{max}} \left(\displaystyle\sum_{i \in F_t} r_{i,k}\right) \leq R_k \label{eq:f4} \\
	& \;\qquad \forall k \in \{0, \dots, M-1\} \notag \\
	& \qquad F_t = \{i \in V | s_i \leq t < s_i+d_i\} \notag
\end{align}

The objective of the RCPSP is to find a feasible schedule $W$ with the minimal schedule length $C_{max}$.
The schedule length is the latest finish time of any activity (Equations \eqref{eq:f1},\eqref{eq:f2}).
Equation \eqref{eq:f3} ensures that all precedence relations are satisfied.
A schedule is feasible if all precedence relations are satisfied and the resources are not overloaded,
i.e. the activities requirements do not exceed the capacity of any resource at any time (Equation~\eqref{eq:f4}).

\begin{table}[th]
	\center
	\scalebox{0.85}{
	\begin{tabular}{*5{|c}|} \hline
		\textbf{Activity} $i$ & $d_i$ & $r_{i,0}$ & $r_{i,1}$ & \textbf{Successors}\\ \hline
		0 & 0 & 0 & 0 & $\{1,2\}$ \\ \hline
		1 & 4 & 5 & 3 & $\{3,6\}$ \\ \hline
		2 & 3 & 2 & 1 & $\{4,5\}$ \\ \hline
		3 & 5 & 3 & 2 & $\{5,10\}$ \\ \hline
		4 & 5 & 2 & 3 & $\{7\}$ \\ \hline
		5 & 3 & 3 & 4 & $\{8,9\}$ \\ \hline
		6 & 2 & 4 & 1 & $\{7,9\}$ \\ \hline
		7 & 4 & 2 & 2 & $\{8,10\}$ \\ \hline
		8 & 2 & 4 & 5 & $\{11\}$ \\ \hline
		9 & 3 & 1 & 2 & $\{11\}$ \\ \hline
		10 & 4 & 2 & 2 & $\{11\}$ \\ \hline
		11 & 0 & 0 & 0 & $\{\}$ \\ \hline
	\end{tabular}}
	\caption{Data of an example instance.}
	\label{tbl:exampleInstance}
\end{table}

\subsection{Instance Example}

\begin{figure}[bh]
	\begin{center}
		\includegraphics[width=0.5\textwidth]{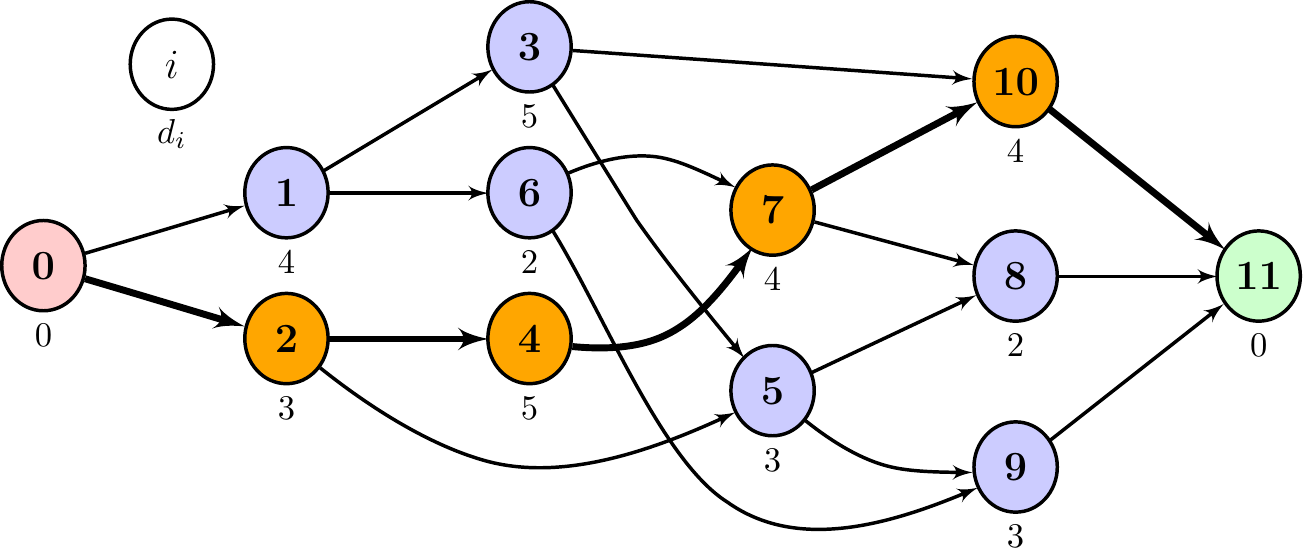}
	\end{center}
	\caption{Graph of precedences for the example instance.}
	\label{fig:exampleGraph}
\end{figure}

\begin{figure*}[t]
	\centering
	\includegraphics[trim = 2mm 0mm 0mm 0mm, clip, width=0.7\textwidth]{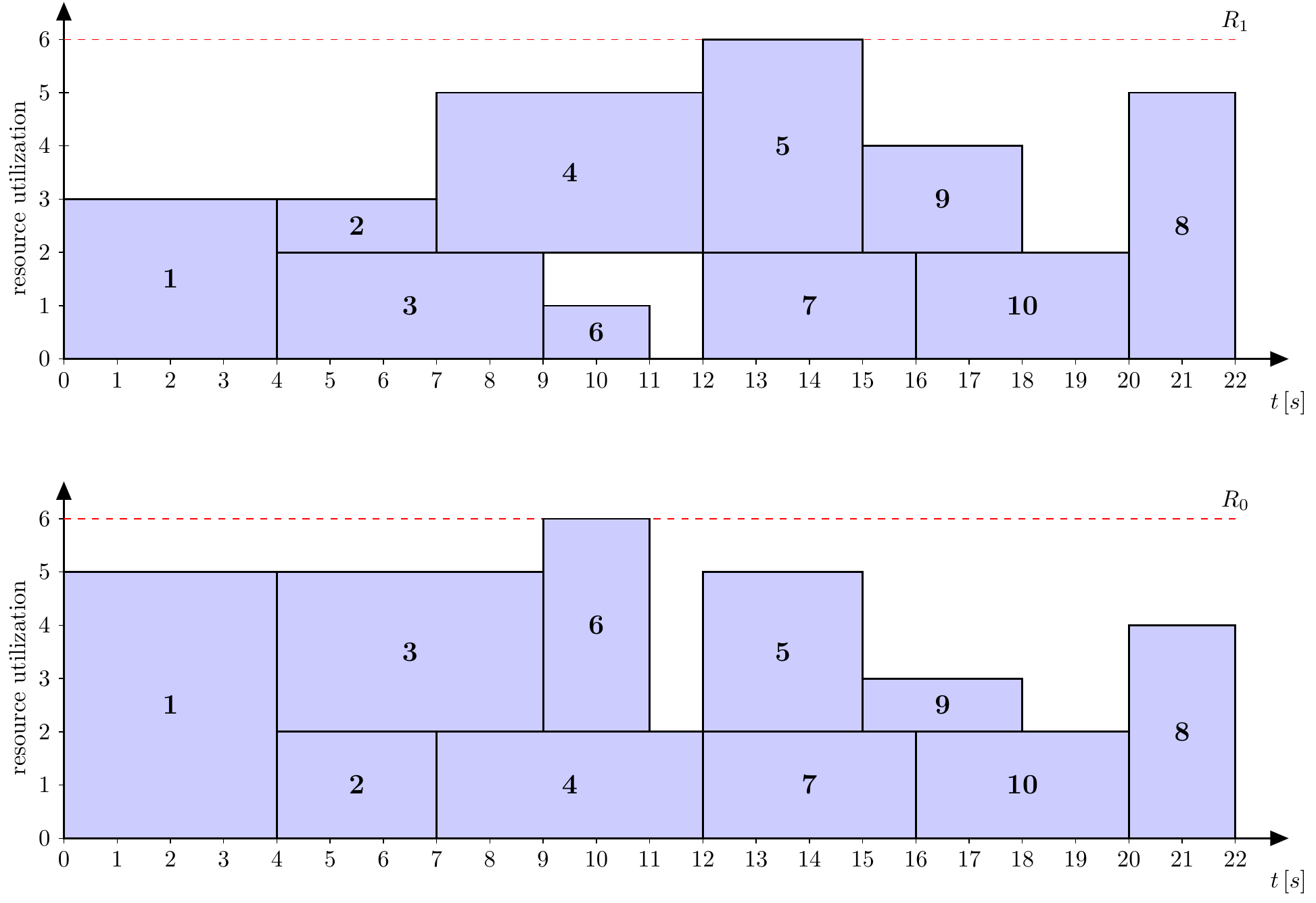}
	\caption{Utilization of resources for the example instance.}
	\label{fig:exampleSource}
\end{figure*}

The data of an example instance are showed in Table~\ref{tbl:exampleInstance}.
In the project there are 10 non-dummy activities and 2 renewable resources with maximal capacity 6.
The corresponding graph of precedences is shown in Figure~\ref{fig:exampleGraph}.
The critical path is highlighted by bold lines and its length is 16.
One of the feasible solutions of the instance is the activity order $W = \{0,1,2,3,4,6,5,7,9,10,8,11\}$ with $C_{max} = 22$.
The resource utilization for this order is depicted in Figure~\ref{fig:exampleSource}.

\section{Brief description of the Tabu Search meta-heuristic}\label{sec:briefTSDesc}

To move through solution space $\mathcal{W}$, a transformation of the current solution to a neighborhood solution is required.
This transformation is called a \emph{move} which can be seen as a light solution modification like a swap of two elements in an order, etc.

The Tabu Search meta-heuristic was proposed by Glover \cite{Glover1986} as an improvement of the \emph{local search} technique \cite{TS_INTRO}.
A local search algorithm starts from the \emph{initial solution} and iteratively improves this solution by applying the best neighborhood moves until a local optimum is reached, 
whereas the Tabu Search introduces a short-term memory called \emph{Tabu List} which reduces the probability of getting stuck in a local optimum or plateau by forbidding the previously visited solutions.
As a consequence, not only improving solutions are permitted and the search process is able to climb to the hills in the search space $\mathcal{W}$ if it is necessary.

Due to efficiency, the Tabu List usually contains only parts of solutions or several previously applied \emph{moves}.
As moves or parts of solutions do not have to be unique, it is possible that a forbidden move leads to the best solution.
In this case it is reasonable to permit the move since the resulting solution was not visited before.
In general, exceptions allowing forbidden moves are called \emph{aspiration criteria} \cite{TS_INTRO}.

The quality of the resulting solutions can be further improved by a suitable search strategy.
For example, if a current location in the solution space is promising, i.e. the best solution was found recently,
then a more thorough search is performed -- \emph{intensification}.
It can be accomplished by concentrating more computational power to this locality of the space.
Opposite to that, if a current location is unpromising, i.e. only poor solutions were found, then \emph{diversification} is performed.
The diversification moves the current search location to another one where better solutions could be found.
It is often realized by applying a few random moves.

The Tabu Search process is stopped if a stop criterion is met.
The stop criterion can be the number of iterations, achieved quality of the best found solution, the maximal number of iterations since the last best solution was found, etc.

\section{Exploration of the Solution Space}\label{sec:PTSG}

\subsection{Creating Initial Activity Order\label{sec:initialOrder}}

Our Tabu Search algorithm starts from initial feasible solution $W^{init} \in \mathcal{W}$ which is created in the following way:
First of all, the longest paths in graph $G$ from the start activity $0$ to all other activities are found.
The weight of each graph edge $(i,j) \in E$ is set to 1.
Activities with the same maximal distance from the start activity are grouped to \emph{levels}.
The level $l_k$ corresponds to all activities with maximal distance $k$ from the start activity,
therefore, activity $0$ is at level $l_0$ and activity $N-1$ is at level $l_{max}$ 
where subscript \emph{max} corresponds with the last level number.
The final feasible schedule can be created from levels such that $W = \{\{l_0\}, \dots,\{l_{max}\}\}$.
Alternative feasible schedules can be created by shuffling the activities on the same level.

\subsection{Move Transformation}\label{sec:moveTR}

\begin{figure}[hb]
	\centering
	\includegraphics{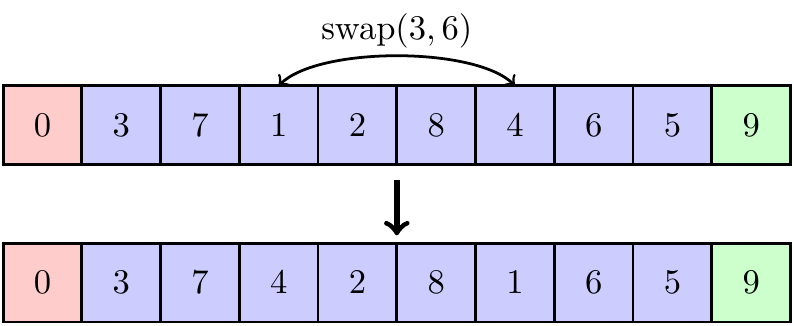}
	\caption{Example of the swap move.}
	\label{fig:moveTypes}
\end{figure}

Schedule order \emph{W} is changed in our Tabu Search algorithm by a \emph{swap} move.
A simple example is illustrated in Figure \ref{fig:moveTypes}. 
Two dummy activities ($0$ and $N-1$) cannot be swapped due to precedence constraints,
therefore the activity at $w_0 $ is always $0$ and the activity at $w_{N-1}$ is always $N-1$.
The swap move is defined as $\textrm{swap}(u,v)$ where $u$ and $v$ are swapped indices.
As $\textrm{swap}(u,v)$ modifies a schedule in the same way as $\textrm{swap}(v,u)$ only swaps with $u < v$ are taken into account without loss of generality.

Let $W \in \mathcal{W}$ be a feasible activity order.
A feasible order means that there is not a violated precedence relation. 
A \emph{feasible move} is a move which does not violate any precedence relation,
therefore if this move is applied to a feasible schedule then the modified schedule will be feasible as well.
Move $\textrm{swap}(u,v)$ is feasible if the following equations are satisfied.

\noindent
\begin{gather}
	(w_u, w_x) \notin E \qquad \forall x \in \{u+1, \dots, v\}\label{eq:breakI} \\
 	(w_x, w_v) \notin E \qquad \forall x \in \{u, \dots, v-1\}\label{eq:breakJ}
\end{gather}

The First Equation \eqref{eq:breakI} means that there are no edges from activity $w_u$ to the activities at indices from $u+1$ to $v$.
If there is any edge, then activity $w_u$ cannot be moved to position $v$ without a precedence violation.
In a similar way, Equation~\eqref{eq:breakJ} states that activity $w_v$ cannot be moved to index $u$ if there is a precedence relation that becomes violated.

\subsection{Neighborhood Generation}\label{sec:neigh}

\emph{Full neighborhood} $\mathcal{N}_{full}\left(W\right) \subseteq \mathcal{W}$ of schedule $W$ is a set of schedules obtained by applying all feasible moves.
Since the full neighborhood is usually too large to be evaluated in a reasonable time only a subset of the neighborhood is usually taken into account.
Such a subset will be called as a \emph{reduced neighborhood} denoted $\mathcal{N}_{reduced}\left(W\right)$.
In the reduced neighborhood, moves are restricted to all $\mathrm{swap}(u,v)$, where $u < v$ and $\left|v-u\right| \leq \delta$.
Value $\delta$ is the maximal distance between the swapped activities in order~$W$.
The size of the neighborhood $\left| \mathcal{N}_{reduced}\left(W\right) \right|$ is parametrized by $\delta$. 

There are two reasons why only feasible moves are applied.
The neighborhood size is reduced without noticeable deterioration of the project makespan and there is no need to check the feasibility of schedules.

\subsection{Filtering Infeasible Moves}\label{sec:filter}

In order to saturate a GPU, feasible schedules in $\mathcal{N}_{reduced}$ should be evaluated in a parallel way by dividing the schedules equally among the threads.
Since the evaluation of a schedule is much more time-consuming than checking whether a move is feasible
it is advantageous to filter out all infeasible moves before the neighborhood evaluation.
It reduces the branch divergency of warps, hence the overall performance of the resources evaluation is improved.

In Algorithm~\ref{alg:filter} is shown how infeasible moves are filtered out from the neighborhood.
The filter works in two phases since it was discovered that it is more effective due to the lower branch divergency than to filter out all the infeasible moves at once.
In the end, only part of the array with feasible moves is taken into account in the neighborhood evaluation.

\begin{algorithm}[ht]
	\linespread{1}{
	\begin{algorithmic}[1]
		\REQUIRE $\mathcal{N}_{reduced}(W)$
		\ENSURE It filters out infeasible moves from the reduced neighborhood.

		\STATE Let $\mathit{MovesArray}$ be an array containing all potential swaps in $\mathcal{N}_{reduced}(W)$.
		\STATE All moves not satisfying Equation \eqref{eq:breakI} are removed, i.e. set empty.
		\STATE Reorder $\mathit{MovesArray}$ such that all empty moves are in the end of the array.
		\STATE Remove moves that do not satisfy Equation \eqref{eq:breakJ}.
		\STATE Move all feasible moves to the beginning of $\mathit{MovesArray}$.
	\end{algorithmic}}
	\caption{Removing infeasible moves from the reduced neighborhood.}
	\label{alg:filter}
\end{algorithm}

\subsection{Simple Tabu List and Cache}\label{sec:SimpleTL}

The tabu list in \cite{FlowShopGPU} is not suitable for a GPU
since it is necessary to go through all moves in the tabu list to decide whether a move is in the tabu list or not.
As a consequence it places higher demands on the device memory bandwidth.
To avoid this a simple and efficient Simple Tabu List (STL) with constant algorithmic complexity is proposed.
Access to the STL is performed like an access to a circular buffer.
Its size is fixed and is equal to $\left|tabuList\right|$.
In our case, the STL stores \emph{swap} moves.
Each $\mathrm{swap}(u, v)$ is stored to the STL as a pair of swapped indices $(u,v)$.
A special value is used for an empty move, e.g. $\mathrm{swap}(0,0)$.
At each iteration of the TS algorithm one move is added (see Algorithm \ref{alg:addMoveSTL})
and the oldest one is removed if the STL is full.
 
\begin{algorithm}[hb]
	\center
	\linespread{1}{
	\begin{algorithmic}[1]
		\REQUIRE $\mathit{tabuCache} - \textrm{STL cache.}$
		\REQUIRE $(u,v) - \textrm{Swap move indices.}$
		\ENSURE It returns true if the move is in the STL, otherwise false.

		\RETURN $\mathit{tabuCache}$[$u,v$]
	\end{algorithmic}}
	\caption{Check if a move is in the STL.}
	\label{alg:CheckMoveSTL}
\end{algorithm}

The \emph{Tabu Cache} (TC) was proposed for effective checking if the move is in the STL.
It is illustrated in Algorithm~\ref{alg:CheckMoveSTL}.
Checking if a swap is in the STL occurs much more often than adding a new move
since a move is added only once per iteration and check if the move is in the STL occurs for every neighborhood schedule.
The TC is implemented as a 2-dimensional $N \times N$ boolean array which is synchronized with the STL.
A check if a move is in the STL requires one read operation, thus, the required memory bandwidth is very low.
It is obvious that a check if the move is in the STL has $O(1)$ algorithmic complexity.

\begin{algorithm}[ht]
	\center
	\linespread{1}{
	\begin{algorithmic}[1]
		\REQUIRE $\mathit{tabuList} - \textrm{Fixed size array.}$
		\REQUIRE $\mathit{tabuCache} - \textrm{STL cache.}$
		\REQUIRE $\mathit{writeIndex} - \textrm{Current write position.}$
		\REQUIRE $(u,v) - \textrm{Swap move indices.}$
		\ENSURE Add move to STL and update TC.

		\STATE $(u_{old},v_{old}) = \mathit{tabuList}$[$\mathit{writeIndex}$]
		\STATE $\mathit{tabuCache}$[$u_{old},v_{old}$] = \textbf{false}
		\STATE $\mathit{tabuList}$[$\mathit{writeIndex}$]$ = (u,v)$
		\STATE $\mathit{tabuCache}$[$u,v$] = \textbf{true}
		\STATE $\mathit{writeIndex} = (\mathit{writeIndex}+1)\,\%\,\left|tabuList\right|$
	\end{algorithmic}}
	\caption{Add a move to the STL.}
	\label{alg:addMoveSTL}
\end{algorithm}

\section{Schedule Evaluation}\label{sec:scheduleEvaluation}

During the evaluation of $W$, precedence relations and resource constraints have to be taken into account to calculate activities start times $s_i$ and $C_{max}$.
The precedence earliest start time $es_i^{\mathit{prec}}$ of activity $i$ can be calculated as $\max_{\forall (j,i) \in E} \left(s_j+d_j\right)$, where $j$ are predecessors of activity $i$.
The resources earliest start time $es_i^{\mathit{res}}$ can be computed using either a \emph{time-indexed} or \emph{capacity-indexed} resources evaluation algorithm.
The capacity-indexed algorithm is a completely new approach to the best of our knowledge, whereas the time-indexed algorithm is well-known \cite{KelleySGS}.
The names of the algorithms were selected with respect to the indexed unit of a resource state array.
According to a heuristic the probable faster resources evaluation algorithm is selected in the schedule evaluation procedure.
Having considered both precedence and resource constraints the final earliest start time is $es_i = \max(es_i^{\mathit{prec}}, es_i^{\mathit{res}})$.

\subsection{Capacity-indexed resources evaluation}

\subsubsection{Required Data Structures}

The most difficult part during the project makespan evaluation
is computation of the activities' start times with respect to the resource capacities.
In our approach, the evaluation of resources requires one array $c_k$ with length $R_k$ per resource~$k$.
Value $c_k[R_k-r_{i,k}]$ corresponds to the earliest resource start time of activity $i$
with resource requirement $r_{i,k} > 0$ on resource~$k$.
At the start of the evaluation, all the resources arrays are set to zeros.
After that, activities are added one by one to a schedule according to $W$ and arrays are updated with respect to the activity requirements and precedences.
The resources arrays are ordered descendly, i.e. $c_k[R_k-l] \leq c_k[R_k-l-1]\;|\;\forall l \in \{1,\dots,R_k-1\}$.
The state of resources is represented as a set $C = \{c_0, \dots, c_{M-1}\}$.

\subsubsection{The Earliest Resources Start Time}

Resource earliest start time $es_i^{\mathit{res}} \in \mathbb{N}$ of activity $i$
with respect to an occupation of resources can be calculated using Equation~\eqref{equ:est_capacity}.
It is guaranteed that resources are not overloaded if activity $i$ start time $s_i \geq es_i^{\mathit{res}}$.
Final activity start time $s_i$ can be more delayed due to the precedence relations.

\begin{equation}
	es_i^{\mathit{res}} = \begin{cases}
		\displaystyle \max_{k \in \{0,\dots,M-1\}:\;r_{i,k} > 0\hspace{-0.15cm}} \hspace{-1cm} c_k[R_k-r_{i,k}] & \exists r_{i,k} > 0 \\
		0 & \text{otherwise}
	\end{cases}
	\label{equ:est_capacity}
\end{equation}

\vspace{0.5em}

\subsubsection{Update of the Resources Arrays}

If activity $i$ is added into the schedule, resources arrays $C$ have to be updated by Algorithm~\ref{alg:resourcesAddActivity}.
\begin{algorithm}[!hb]
\linespread{1}{
\begin{algorithmic}[1]
\REQUIRE $r_{i,k},d_i,C,\mathcal{R}$
\REQUIRE $\mathit{copy} - \textrm{Auxiliary array with length}\;R_{max}\textrm{)}.$
\REQUIRE $s_i - \textrm{Scheduled start time of activity}\;i.$
\ENSURE Update $C$ - activity $i$ is added.

\FOR{($k$ = 0; $k < M$; ++$k$)}
  \STATE $\mathit{requiredEffort} = r_{i,k} \cdot d_i$
  \IF{($\mathit{requiredEffort} > 0$)}
    \STATE $\mathit{resIdx} = 0;\;\mathit{copyIdx} = 0$
    \STATE $\mathit{newTime} = s_i+d_i$
    \WHILE{($\mathit{requiredEffort} > 0$ \textrm{AND} $\mathit{resIdx} < R_k$)}
      \IF{($c_k$[$\mathit{resIdx}$]$\;< \mathit{newTime}$)}
	\IF{($\mathit{copyIdx} \geq r_{i,k}$)}
	  \STATE $\mathit{newTime} = \mathit{copy}$[$\mathit{copyIdx}-r_{i,k}$]
	\ENDIF
	\STATE $\mathit{timeDiff} = \mathit{newTime}-\max(c_k$[$\mathit{resIdx}$]$, s_i)$
	\IF{($\mathit{requiredEffort}-\mathit{timeDiff} > 0$)}
	  \STATE $\mathit{requiredEffort}$ -{}= $\mathit{timeDiff}$
	  \STATE $\mathit{copy}$[$\mathit{copyIdx}$++]$\;= c_k$[$\mathit{resIdx}$]
	  \STATE $c_k$[$\mathit{resIdx}$]$\;= \mathit{newTime}$
	\ELSE
	  \STATE $c_k$[$\mathit{resIdx}$]$\;= \max(c_k$[$\mathit{resIdx}$]$, s_i)$
	  \STATE $c_k$[$\mathit{resIdx}$] +{}= $\mathit{requiredEffort}$
	  \STATE $\mathit{requiredEffort} = 0$
	\ENDIF
      \ENDIF
      \STATE $\mathit{resIdx} = \mathit{resIdx}+1$
    \ENDWHILE
  \ENDIF
\ENDFOR
\end{algorithmic}}
\caption{Method updates state of resources after adding activity $i$.}
\label{alg:resourcesAddActivity}
\end{algorithm}
Each resource array $c_k$ is updated individually (line~1).
Value $\mathit{requiredEffort} = r_{i,k} \cdot d_i$ will be called \emph{Required Resource Effort}.
Activity $i$ can be added to the schedule if and only if each resource is able to provide its Required Resource Effort.
In other words, variable \emph{requiredEffort} has to be decremented to zero (lines 2, 13, 19) for each resource $k$.

It is performed by setting $r_{i,k}$ elements of $c_k$ to the activity finish time $s_i+d_i$ after the last $c_k \geq s_i+d_i$.
Until the variable \emph{requiredEffort} is zero, the shifted (right shift about $r_{i,k}$) copy of the original resource array $c_k$
(original values are stored in $\mathit{copy}$ auxiliary variable) is made.
The complexity of the algorithm is $O(M \cdot R_{max})$.

\begin{figure}[!hb]
	\begin{center}
		\includegraphics{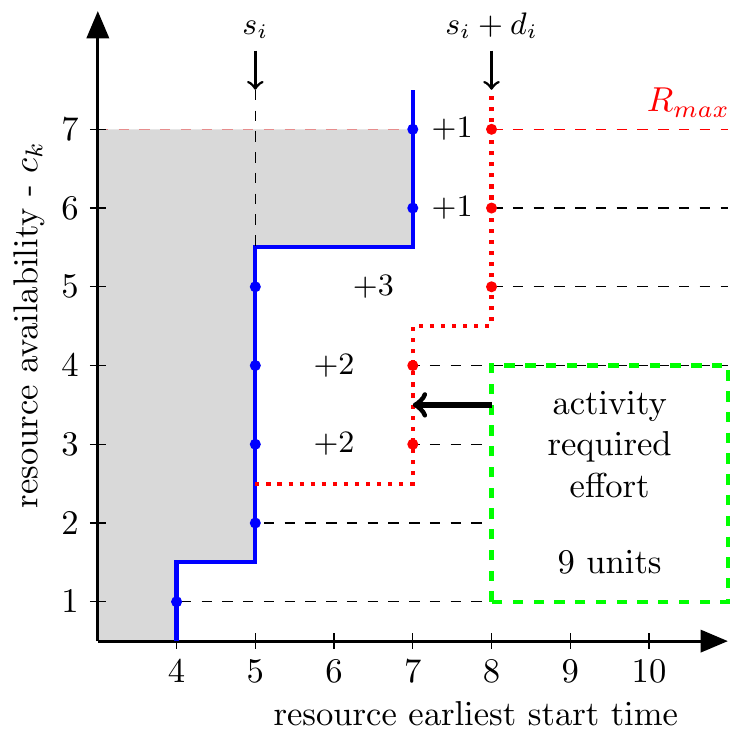}
	\end{center}
	\caption{An example of the resource state update.}
	\label{fig:exampleUpdateResource}
\end{figure}

The algorithm is illustrated on an example in Figure \ref{fig:exampleUpdateResource}.
There is one resource with maximal capacity 7.
Added activity $i$ requires 3 resource units (i.e. $r_{i,k} = 3$) and its duration $d_i$ is 3.
The activity was scheduled at $s_i = 5$.
The solid line corresponds to the original resource state $c_k = \{7, 7, 5, 5, 5, 5, 4\}$
and the dotted line corresponds to the updated resource $c_k' = \{8, 8, 8, 7, 7, 5, 4\}$.
Activity required effort is depicted by a square with dashed border.
The positive numbers between solid and dotted lines are effort contributions
when old start time (solid line) will be changed to the new start time (dotted line).
It should be noticed, that the sum of all contributions is the $\mathit{requiredEffort}$ for a given activity.

\subsection{Time-indexed resources evaluation}

\subsubsection{Required Data Structures}

In the time-indexed evaluation algorithm the state of each resource $k$ is stored in array $\tau_k$.
Each element $\tau_k[t]$ corresponds to the number of available resource units that resource $k$ is able to provide at time $t \in \{0,\dots,\mathrm{UB}_{C_{max}}\}$,
where $\mathrm{UB}_{C_{max}}$ is the upper bound of the makespan which can be calculated as e.g. $\sum_{\forall i \in V} d_i$.
Each $\tau_k$ array has initialized all its elements to the $R_k$ value before the start of the evaluation.
The state of all resources will be denoted as $T = \{\tau_0,\dots,\tau_{M-1}\}$.

\subsubsection{The Earliest Resources Start Time}

\begin{algorithm}[bh]
	\linespread{1}{
	\begin{algorithmic}[1]
		\REQUIRE $r_{i,k}, d_i, \mathcal{R}, \mathrm{UB}_{C_{max}}, T$
		\REQUIRE $es_i^{\mathit{prec}} - \textrm{The precedence earliest start time.}$
		\ENSURE Calculate the earliest start time $es_i$ of activity $i$.

		\STATE $\mathit{loadTime} = 0$;
		\FOR{($t$ = $es_i^{\mathit{prec}}$; $t < \mathrm{UB}_{C_{max}}$ \textbf{AND} $\mathit{loadTime} < d_i$; ++$t$)}
			\STATE $\mathit{sufficientCapacity} = \mathrm{true}$
			\FOR{($k$ = $0$; $k < M-1$; ++$k$)}
				\IF{($\tau_k$[$t$] $<$ $r_{i,k}$)}
					\STATE $\mathit{loadTime} = 0$
					\STATE $\mathit{sufficientCapacity} = \mathrm{false}$
				\ENDIF
			\ENDFOR
			\IF{($\mathit{sufficientCapacity} == \mathrm{true}$)}
				\STATE ++$\mathit{loadTime}$
			\ENDIF
		\ENDFOR
		\RETURN $t-\mathit{loadTime}$
	\end{algorithmic}}
	\caption{Algorithm calculates the earliest start time of activity $i$.}
	\label{alg:est}
\end{algorithm}

The earliest start time of activity $i$ can be calculated using Algorithm~\ref{alg:est}.
In the algorithm, the $\mathit{loadTime}$ variable corresponds with the number of consecutive time units in which resources are able to meet resource requirements of activity $i$.
If $\mathit{loadTime} = d_i$ then a time interval into which activity $i$ can be scheduled was found.
Having considered variable $t$ as a finish time of a candidate interval, the resulting interval is the first interval $[t-\mathit{loadTime},t) \cap \mathbb{N}$ such that $\mathit{loadTime} = d_i$.
The final earliest start time is the lower endpoint of the interval.

\subsubsection{Update of the Resources Arrays}

The state of resources is updated as is shown in Algorithm~\ref{alg:time_upd}.
Having scheduled activity $i$ at $s_i$ the $\tau_k$ arrays have to be updated in the $[s_i,s_i+d_i)$ interval.
For each resource $k$, values in the interval are decreased by $r_{i,k}$ units.

\begin{algorithm}[ht]
	\linespread{1}{
	\begin{algorithmic}
		\REQUIRE $d_i, T, r_{i,k}$
		\REQUIRE $s_i - \textrm{Scheduled start time of activity}\;i$.
		\ENSURE It updates state of resources $T$.

		\FOR{($k = 0$; $k < M-1$; ++$k$)}
			\FOR{($t = s_i$; $t < s_i+d_i$; ++$t$)}
				\STATE $\tau_k[t]$ -{}= $r_{i,k}$
			\ENDFOR
		\ENDFOR
	\end{algorithmic}}
	\caption{Updating of resources after adding activity~$i$.}
	\label{alg:time_upd}
\end{algorithm}

\subsection{Schedule Evaluation Procedure}\label{sec:completeEvaluation}

The schedule evaluation procedure is shown in Algorithm~\ref{alg:scheduleEvaluate}.
Activities are read one by one from the activities order $W$.
For each activity $w_u$, all its predecessors are found and the precedence relations 
are used to update activity $w_u$'s precedence earliest start time $es_{w_u}^{\mathit{prec}} \in \mathbb{N}$ (see lines 3--6).
Then the resources restrictions are checked and the start time is adjusted to $s_{w_u} = \max (es_{w_u}^{\mathit{prec}},es_{w_u}^{\mathit{res}})$.
\begin{algorithm}[bh]
\linespread{1}{
\begin{algorithmic}[1]
\REQUIRE $W,C,T,E$
\ENSURE Calculate $C_{max}$ and the activities' start times.

\STATE $C_{max} = 0$
\FOR{($u = 0$; $u < N$; ++$u$)}
  \STATE $es_{w_u}^{\mathit{prec}} = 0$
  \FORALL{$\left((j,w_u) \in E\right)$}
      \STATE $es_{w_u}^{\mathit{prec}} = \max (es_{w_u}^{\mathit{prec}}, s_j + d_j)$
  \ENDFOR
  \STATE $es_{w_u}^{\mathit{res}} = \textrm{getEarliestResourcesTime}(\textrm{activity}\;w_u, es_{w_u}^{prec})$\hspace*{-1mm}
  \STATE $s_{w_u} = \max(es_{w_u}^{\mathit{prec}}, es_{w_u}^{\mathit{res}})$
  \STATE $\textrm{updateResources}(\textrm{activity}\;w_u, s_{w_u})$
  \STATE Mark current activity $w_u$ as scheduled.
  \STATE $C_{max} = \max (C_{max}, s_{w_u}+d_{w_u})$
\ENDFOR
\RETURN $C_{max}$
\end{algorithmic}}
\caption{Complete schedule evaluation.}
\label{alg:scheduleEvaluate}
\end{algorithm}
Project makespan $C_{max}$ is the finish time of activity $N-1$.
As only feasible moves are allowed, an infeasibility test of the resulting schedules is not required.

\subsection{Heuristic Selection of Resources Evaluation Algorithms}\label{sec:heurEval}

Before the search is started on the GPU, the probable faster resources evaluation algorithm is heuristically selected by decision rules and the required resources arrays are allocated.
To create the rules, the JRip classifier from the Weka data-mining tool \cite{Weka} was learned using pre-calculated attributes shown in Table~\ref{tab:attr}.

\begin{table}[th]
	\centering
	\begin{tabular}{|l|c|} \hline
		Min. resource capacity: & $\displaystyle \min_{\mathclap{\forall k \in \{0,\dots,M-1\}}} R_k$ \\ \hline
		Avg. resource capacity: & $\frac{1}{M} \displaystyle \sum_{\mathclap{\forall k \in \{0,\dots,M-1\}}} R_k$ \\ \hline
		Max. resource capacity: & $\displaystyle \max_{\mathclap{\forall k \in \{0,\dots,M-1\}}} R_k$ \\ \hline
		Avg. activity duration: & $\frac{1}{N} \displaystyle \sum_{\mathclap{\forall i \in V}} d_i$ \\ \hline
		Avg. branch factor: & $\frac{\left|E\right|}{N}$ \\ \hline
		Critical path length: & see Section \ref{sec:problemStatement} \\\hline
		Evaluation algorithm: & CAPACITY/TIME \\ \hline
	\end{tabular}
	\caption{Attributes used for learning.}
	\label{tab:attr}
\end{table}

\begin{figure}[th]
	\centering
	\includegraphics[width=0.5\textwidth]{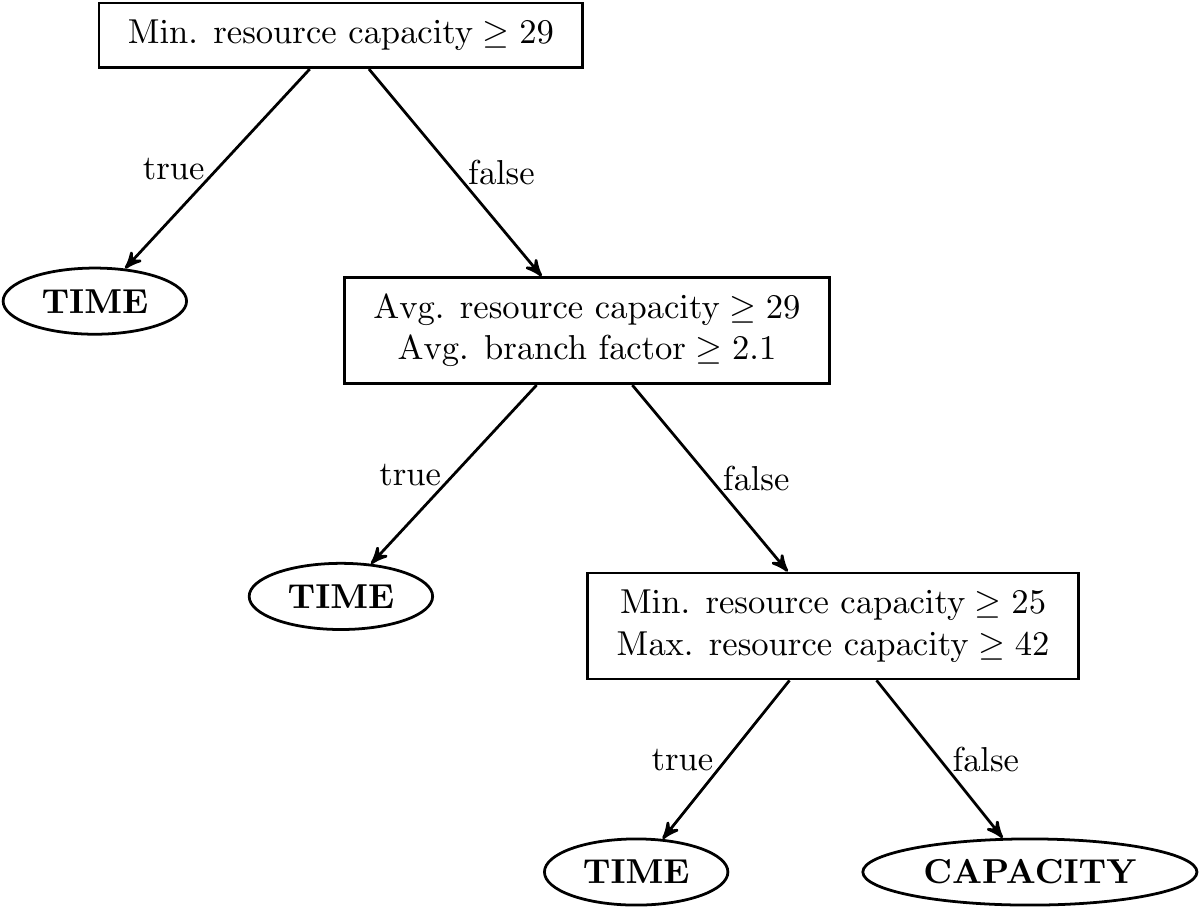}
	\caption{An example of the decision tree.}
	\label{fig:decTree}
\end{figure}

Attribute ``Evaluation algorithm'' determines the class, i.e. the time-indexed or capacity-indexed evaluation algorithm, to which the classifier should classify.
As the final class dependends on the hardware and instance parameters it is necessary to determine the probable correct class by measuring
--- each evaluation algorithm was selected for a small number of iterations and the faster one was selected as the desired one.
The resulting rules heuristically decide which of the two algorithms should be more effective for a given instance.
The rules can be transformed into a decision tree as is shown in Figure~\ref{fig:decTree}.
Once the rules are created they can be applied to other similar instances without any measuring overhead as well.
To show the effectivity and usefulness of the heuristic the experiments were performed in Section~\ref{sec:ExRes}.

\section{Parallel Tabu Search for CUDA platform}\label{sec:PTS_CUDA}

Our Parallel Tabu Search for GPU (PTSG) is proposed with respect to the maximal degree of parallelization since thousands of CUDA threads are required to be fully loaded to exploit the graphics card power.
In our approach, the parallelization is carried out in two ways.
The first one is a parallelization performed within the scope of a block, for example the parallel filter (see Section~\ref{sec:filter}), the parallel neighborhood evaluation, and other parallel reductions.
The second one is a parallelization introduced by launching many blocks on the multi-processors simultaneously.

\begin{figure}[th]
	\centering
	\includegraphics[width=0.48\textwidth]{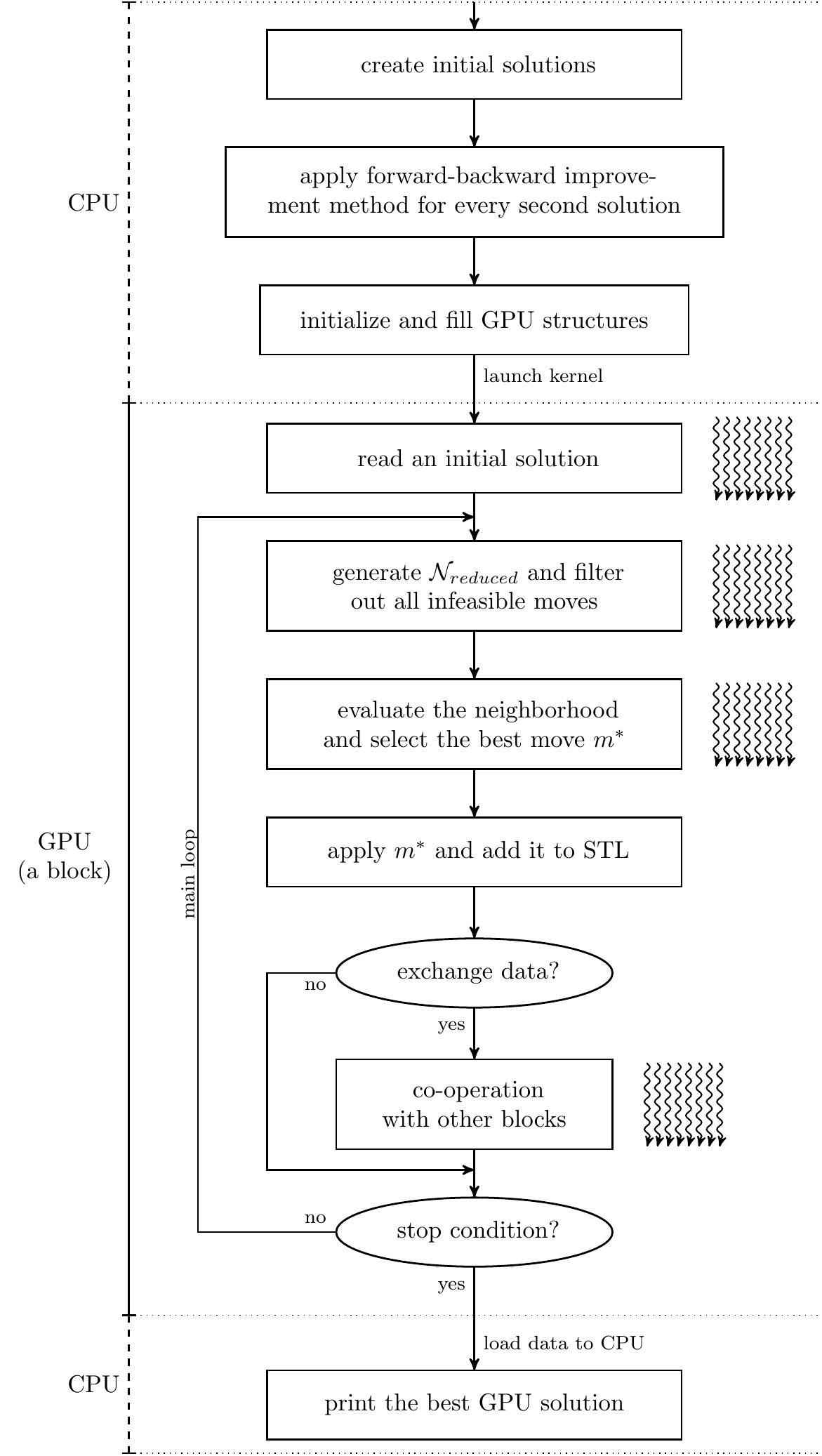}
	\caption{Parallel Tabu Search for the CUDA platform.}
	\label{fig:PTSprocessflow}
\end{figure}

The basic steps of the PTSG are described in Figure~\ref{fig:PTSprocessflow}.
First of all, an instance is read and the initial solutions are created in accordance with Section~\ref{sec:initialOrder}.
After that, every second solution is improved by using the forward-backward improvement method.
The method is iteratively shaking a schedule from the left to the right in order to make a resource profile straight as long as the schedule is getting shorter.
To get more details about the method, refer to the original article by Li and Willis~\cite{Li1992370}.
The created solutions are copied into a \emph{working set}, i.e. a set of shared solutions.
The best solution in the working set will be called the \emph{global best solution} and its makespan will be denoted as $C_{max}^*$.
Furthermore, the block's Tabu Lists and Tabu Caches are initialized and auxiliary arrays such as $\tau_k$, and $c_k$ are allocated.
Having had prepared required data-structures, the host is ready to launch the kernel.

In the GPU part, every block is an independent Tabu Search instance communicating with the others through the global memory (see Section~\ref{sec:cooperation}).
At the beginning, every block reads an initial solution from the working set.
After that, the search is started for a specified number of iterations of the main loop.
In the main loop, the neighborhood is generated, evaluated and the best move $m^*$ is selected, applied and added into the STL.
Move $m^*$ leads to the best criterion improvement or to the smallest criterion deterioration.
This move cannot be in the STL with one exception --- the move leads to the global best solution.
At the end of an iteration the solutions are exchanged through the working set if the communication conditions are satisfied (see Section~\ref{sec:cooperation}).
The search is stopped if the specified number of iterations was achieved or $C_{max}^*$ is equal to the length of a critical path.

After the termination of the kernel, the best global solution is copied from the global memory to the host memory.
The solution is printed and all allocated data-structures are freed.

\subsection{Co-operation Among CUDA Blocks and Iterations Distribution}\label{sec:cooperation}

To assure the high quality solutions, the co-operation among Tabu Search instances is accomplished by exchanging solutions through the working set $F$ that has a fixed number of solutions $\left|F\right|$.
Each solution $k \in F$ consists of the order of activities $W^k$, makespan $C_{max}^k$, the tabu list and iterations counter $\mathit{IC}^k$.
The solutions exchange takes place if the last read solution has not been improved for more than $\mathcal{I}_{assigned}$ iterations or the block found an improvement of the last read solution.
The block writes the best found solution to $F$ if it improves the last read solution and reads the next solution from $F$.
Since the working set could be accessed by many blocks at the same time, it is necessary to use read/write locks in order to maintain data integrity.
The co-operation among Tabu Search instances was inspired by James~et~al.~\cite{James2009810}.

After the block has read a solution from the working set, it is checked whether the solution was not read more than $\Phi_{max}$ times without being improved.
If it is the case, a small number $\Phi_{steps}$ of random feasible swaps is applied to randomize the read solution --- diversification.
After that, the read solution $k \in F$ has assigned the number of iterations $\mathcal{I}_{assigned}$ according to the following equation.
\begin{multline}
	\displaystyle \mathcal{I}_{assigned} = \Bigg\lfloor\overbrace{\frac{1}{5}\frac{\mathcal{I}_{block}}{\mathcal{I}_{total}}}^{quantity}
	\Bigg(\overbrace{0.8e^{-100\left(\frac{C_{max}^k}{\vphantom{\mathcal{E^2}}C_{max}^*}-1\right)}}^{quality} \\
	+\overbrace{0.2e^{-4\left(\frac{\mathit{IC}^k}{\vphantom{\mathcal{E^2}}\mathcal{I}_{block}}\right)}}^{intactness}\Bigg)\Bigg\rfloor
	\label{eq:iterBalancing}
\end{multline}

$\mathcal{I}_{block}$ is the number of iterations assigned to each Tabu Search instance and $\mathcal{I}_{total}$
is the total number of iterations calculated as $\mathcal{I}_{block}B$, where $B$ is the number of launched blocks.
The part denoted as \emph{quantity} corresponds to the maximal number of iterations which can be assigned to read solution $k$.
It is ensured, that at least 5 solutions are read from the working set by each block.
The \emph{quality} part takes into account the quality of read solution $k$.
It is obvious that the high quality solutions are preferred to poor ones --- intensification.
And the last part \emph{intactness} guarantees that each solution $k \in F$ has been given some iterations to prove the quality.

\subsection{Memory Model}\label{sec:MemMod}

The placement of the data-structures is a crucial task highly influencing the effectiveness of the GPU program,
therefore, each decision should be considered thoroughly with respect to the access pattern, required bandwidth and data visibility (local or shared data).
In the shared memory current block order $W_{block}$, durations of activities $D$, and auxiliary arrays are stored.
Although $D$ is a read-only array which could be located in the constants memory, it was moved to the shared memory due to the higher bandwidth.
The texture memory is used for storing read-only data as $r_{i,k}$ values and predecessors of the activities.
In the local memory, private data-structures of each thread are located, i.e. resources arrays $c_k$, $\tau_k$, and start times of activities $S$.
The long latency of the memory is compensated by using a partial coalescing since the arrays are often accessed
at the same relative indices as the majority of threads evaluate similar schedules ($W_{block}$ + swap move).
Finally, the global memory is employed to store the working set $F$.

\section{Experimental Results}\label{sec:ExRes}

Experiments were performed on the AMD Phenom(tm) II X4 945 server (4 cores, 8 GB memory) equipped with a mid-range Nvidia Geforce GTX 650 Ti (1\,GB, 768 cuda cores, 4 multiprocessors) graphics card.
The testing environment was the Windows Server 2008 with an installed CUDA toolkit (version 5.0.35) and Microsoft Visual Studio 2010.

The sequential CPU version of the algorithm corresponds to one Tabu Search instance with the exception that solutions are not interchanged ($\left|F\right| = 1$ and $B = 1$).
Instead of using the selection heuristic (see Section \ref{sec:heurEval}) the faster evaluation algorithm was selected dynamically by periodic measuring every 1000 iterations.
The parallel CPU version differs from the sequential version in the neighborhood evaluation.
The feasible schedules in the neighborhood are divided among CPU threads to reduce evaluation time.
Both the CPU and GPU versions were fully optimized with respect to memory access patterns and hardware architecture (cache sizes).
To fully saturate the GPU the maximal number of available registers per CUDA thread was limited to 32
due to possibility to launch 4~blocks on a multiprocessor at once (altogether 16 blocks on the GPU), where each block has 512 CUDA threads.

\begin{table}[hb]
\centering
\scalebox{0.95}{
\begin{tabular}{|*5{c|}} \cline{2-5}
	\multicolumn{1}{c|}{} & \textbf{J30} & \textbf{J60} & \textbf{J90} & \textbf{J120} \\ \hline
	$N$ & 30+2 & 60+2 & 90+2 & 120+2 \\ \hline
	$M$ & \multicolumn{4}{|c|}{4} \\ \hline
	$\left|dataSet\right|$ & 480 & 480 & 480 & 600 \\ \hline
	$\delta$ & 30 & 60 & 60 & 60 \\ \hline
	$\left|tabuList\right|$ & 60 & 250 & 600 & 800 \\ \hline
	$\Phi_{steps}$ & \multicolumn{4}{|c|}{20} \\ \hline
	$\Phi_{max}$ & \multicolumn{4}{|c|}{3} \\ \hline
	$\left|F\right|$ & \multicolumn{4}{|c|}{16} \\ \hline
\end{tabular}}
\caption{PTSG parameters and data-sets information.}
\label{tab:params}
\end{table}

To evaluate the performance and the quality of resulting solutions the benchmark using the well-known J30, J60, J90 and J120 data-sets was performed.
The number of instances in a data-set will be denoted as $\left|dataSet\right|$.
The selected PTSG parameters and data-sets information are stated in Table~\ref{tab:params}.

\begin{table*}[p]
\centering
\scalebox{0.85}{
\begin{tabular}{|l|*6{c|}} \cline{2-7}
\multicolumn{1}{l|}{} & \multicolumn{3}{c|}{\textbf{CPU}} & \multicolumn{3}{c|}{\textbf{GPU}} \\ \hline
$\mathcal{I}_{total}$ & \emph{CPM dev} & \emph{OPT dev} & \emph{Best\_sol} & \emph{CPM dev} & \emph{OPT dev} & \emph{Best\_sol} \\ \hline
10000 & 13.43 & 0.04 & 471 & 13.41 & 0.02 & 473 \\ \hline
20000 & -- & -- & -- & 13.38 & 0.01 & 478 \\ \hline
\end{tabular}}
\caption{Quality of solutions --- J30.}
\label{tab:comparisonQualityJ30}
\end{table*}

\begin{table*}[p]
\centering
\scalebox{0.85}{
\begin{tabular}{|l|*6{c|}} \cline{2-7}
\multicolumn{1}{l|}{} & \multicolumn{3}{c|}{\textbf{CPU}} & \multicolumn{3}{c|}{\textbf{GPU}} \\ \hline
$\mathcal{I}_{total}$ & \emph{CPM dev} & \emph{UB dev} & \emph{Best\_sol} & \emph{CPM dev} & \emph{UB dev} & \emph{Best\_sol} \\ \hline
10000 & 11.13 & 0.51 & 380 & 11.22 & 0.57 & 375 \\ \hline
20000 & -- & -- & -- & 11.08 & 0.47 & 388 \\ \hline
30000 & -- & -- & -- & 10.99 & 0.41 & 394 \\ \hline
50000 & -- & -- & -- & 10.91 & 0.36 & 394 \\ \hline
\end{tabular}}
\caption{Quality of solutions --- J60.}
\label{tab:comparisonQualityJ60}
\end{table*}

\begin{table*}
\centering
\begin{minipage}[b]{0.48\textwidth}
\scalebox{0.8}{
\begin{tabular}{|l|*4{c|}} \cline{2-5}
\multicolumn{1}{l|}{} & $\mathcal{I}_{total}$ & \emph{Comp\_time} & \emph{Sched\_sec} & \emph{Speedup} \\ \hline
CPU seq. & 10000 & 1255 & ~126400 & 1.00 \\ \hline
CPU par. & 10000 & ~343 & ~478300 & 3.65 \\ \hline
GPU & 10000 & ~176 & ~985543 & 7.12 \\ \hline
GPU & 20000 & ~306 & 1120900 & -- \\ \hline
\end{tabular}}
\caption{Performance comparison --- J30.}
\label{tab:comparisonSpeedupJ30}
\end{minipage}\hfill
\begin{minipage}[b]{0.48\textwidth}
\scalebox{0.8}{
\begin{tabular}{|l|*4{c|}} \cline{2-5}
\multicolumn{1}{l|}{} & $\mathcal{I}_{total}$ & \emph{Comp\_time} & \emph{Sched\_sec} & \emph{Speedup} \\ \hline
CPU seq. & 10000 & 7094 & ~~59700 & ~1.00 \\ \hline
CPU par. & 10000 & 1732 & ~248700 & ~4.10 \\ \hline
GPU & 10000 & ~257 & 1733800 & 27.60 \\ \hline
GPU & 20000 & ~485 & 1818600 & -- \\ \hline
GPU & 30000 & ~709 & 1861900 & -- \\ \hline
GPU & 50000 & 1164 & 1879400 & -- \\ \hline
\end{tabular}}
\caption{Performance comparison --- J60.}
\label{tab:comparisonSpeedupJ60}
\end{minipage}
\end{table*}

\begin{table*}[p]
\centering
\scalebox{0.85}{
\begin{tabular}{|l|*6{c|}} \cline{2-7}
\multicolumn{1}{l|}{} & \multicolumn{3}{c|}{\textbf{CPU}} & \multicolumn{3}{c|}{\textbf{GPU}} \\ \hline
$\mathcal{I}_{total}$ & \emph{CPM dev} & \emph{UB dev} & \emph{Best\_sol} & \emph{CPM dev} & \emph{UB dev} & \emph{Best\_sol} \\ \hline
10000 & 10.81 & 0.92 & 367 & 11.04 & 1.09 & 365 \\ \hline
20000 & -- & -- & -- & 10.82 & 0.93 & 371 \\ \hline
30000 & -- & -- & -- & 10.73 & 0.86 & 373 \\ \hline
50000 & -- & -- & -- & 10.56 & 0.74 & 375 \\ \hline
\end{tabular}}
\caption{Quality of solutions --- J90.}
\label{tab:comparisonQualityJ90}
\end{table*}

\begin{table*}[p]
\centering
\scalebox{0.85}{
\begin{tabular}{|l|*6{c|}} \cline{2-7}
\multicolumn{1}{l|}{} & \multicolumn{3}{c|}{\textbf{CPU}} & \multicolumn{3}{c|}{\textbf{GPU}} \\ \hline
$\mathcal{I}_{total}$ & \emph{CPM dev} & \emph{UB dev} & \emph{Best\_sol} & \emph{CPM dev} & \emph{UB dev} & \emph{Best\_sol} \\ \hline
10000 & 33.41 & 2.70 & 215 & 34.67 & 3.50 & 194 \\ \hline
20000 & -- & -- & -- & 34.04 & 3.11 & 208 \\ \hline
30000 & -- & -- & -- & 33.66 & 2.85 & 213 \\ \hline
50000 & -- & -- & -- & 33.54 & 2.76 & 222 \\ \hline
\end{tabular}}
\caption{Quality of solutions --- J120.}
\label{tab:comparisonQualityJ120}
\end{table*}

\begin{table*}
\centering
\begin{minipage}[b]{0.48\textwidth}
\scalebox{0.8}{
\begin{tabular}{|l|*4{c|}} \cline{2-5}
\multicolumn{1}{l|}{} & $\mathcal{I}_{total}$ & \emph{Comp\_time} & \emph{Sched\_sec} & \emph{Speedup} \\ \hline
CPU seq. & 10000 & 20294 & ~~36000 & ~1.00 \\ \hline
CPU par. & 10000 & ~5001 & ~148300 & ~4.06 \\ \hline
GPU & 10000 & ~~475 & 1599600 & 42.70 \\ \hline
GPU & 20000 & ~~923 & 1632000 & -- \\ \hline
GPU & 30000 & ~1348 & 1660700 & -- \\ \hline
GPU & 50000 & ~2221 & 1674000 & -- \\ \hline
\end{tabular}}
\caption{Performance comparison --- J90.}
\label{tab:comparisonSpeedupJ90}
\end{minipage}\hfill
\begin{minipage}[b]{0.48\textwidth}
\scalebox{0.8}{
\begin{tabular}{|l|*4{c|}} \cline{2-5}
\multicolumn{1}{l|}{} & $\mathcal{I}_{total}$ & \emph{Comp\_time} & \emph{Sched\_sec} & \emph{Speedup} \\ \hline
CPU seq. & 10000 & ~148170 & ~~25700 & ~1.00 \\ \hline
CPU par. & 10000 & ~~35812 & ~107200 & ~4.14 \\ \hline
GPU & 10000 & ~~~2938 & 1340400 & 50.40 \\ \hline
GPU & 20000 & ~~~5742 & 1351900 & -- \\ \hline
GPU & 30000 & ~~~8513 & 1353300 & -- \\ \hline
GPU & 50000 & ~~14160 & 1347800 & -- \\ \hline
\end{tabular}}
\caption{Performance comparison --- J120.}
\label{tab:comparisonSpeedupJ120}
\end{minipage}
\end{table*}

\begin{table*}[th]
\centering
\scalebox{0.85}{
\begin{tabular}{|l|*4{c|}} \cline{2-5}
\multicolumn{1}{}{} & \multicolumn{4}{|c|}{\emph{CPM dev}} \\ \hline
\multicolumn{1}{|c|}{\textbf{Algorithm and reference}} & \textbf{J30} & \textbf{J60} & \textbf{J90} & \textbf{J120} \\ \hline
Genetic Algorithm - Gon\c{c}alves et al. \cite{JoseGAFBI} & 13.38 & 10.49 & - & 30.08 \\ \hline
This work - Nvidia Geforce GTX 650 Ti & 13.38 & 10.91 & 10.56 & 33.54 \\ \hline
This work - AMD Phenom(tm) II X4 945 & 13.43 & 11.13 & 10.81 & 33.41 \\ \hline
CARA algorithm - Valls et al. \cite{Valls2003282} & 13.46 & 11.45 & 11.12 & 34.53 \\ \hline
Ant Colony Optimization - Zhou et al. \cite{ACO_RCPSP} & - & 11.42 & - & 35.11 \\ \hline
Tabu Search - Artigues et al. \cite{Artigues2003249} & - & 12.05 & - & 36.16 \\ \hline
Simulated Annealing - Bouleimen and Lecocq \cite{Bouleimen2003268} & - & 11.90 & - & 37.68 \\ \hline
\end{tabular}}
\caption{Comparison with other heuristics.}
\label{tab:comparisonOther}
\end{table*}

The results for the J30 data-set are shown in Tables~\ref{tab:comparisonQualityJ30} and \ref{tab:comparisonSpeedupJ30}.
The \emph{CPM dev} and \emph{OPT dev} values are the average percentage distance from the critical path length and the average percentage distance from the optimal makespan respectively.
\emph{Best\_sol} states the number of optimal solutions which have been proved to be optimal according to the results in the PSPLIB homepage --- \url{http://www.om-db.wi.tum.de/psplib/}.
\emph{Comp\_time} is the total run-time stated in seconds and the \emph{Sched\_sec} is the number of evaluated schedules per second.
It is obvious that the GPU version is able to achieve a similar quality of solutions in terms of \emph{CPM dev} as the CPU version.
Having had $\mathcal{I}_{total}$ doubled, the GPU version found 478 optimal solutions from the 480 solutions in the data-set.
From the performance point of view Table \ref{tab:comparisonSpeedupJ30} reveals a significant improvement in computational time if parallelization is performed.
For example, the parallel CPU version is 3.65 times faster than the sequential CPU version and the GPU is almost 2 times faster than the parallel CPU version.
If $\mathcal{I}_{total}$ is increased to 20000 the GPU is still slightly faster and achieves better quality solutions.

For the J60 data-set the results are shown in Tables \ref{tab:comparisonQualityJ60} and \ref{tab:comparisonSpeedupJ60}, 
where \emph{UB dev} is the average percentage distance from the best currently known upper bounds.
The CPU version gives slightly better solutions for 10000 iterations, but on the other hand if the GPU is given 20000 iterations
the quality of solutions is comparable with the CPU version and the GPU is still 3.56 times faster than the parallel CPU version.
The lower quality of GPU solutions for the same $\mathcal{I}_{total}$ is probably caused by wasting work when many Parallel Tabu Search instances
have read the same solution from the working set and only one writes the best improvement.
It can be noted, that the parallel CPU version is more than 4 times faster than the sequential one.
The reason of that is either better cache utilization or the AMD True Core Scalability technology.

The results in Tables \ref{tab:comparisonQualityJ90} and \ref{tab:comparisonSpeedupJ90} for the J90 data-set show that the GPU is better utilized for bigger instances
and the GPU is more than 10 times faster than the parallel CPU version for the same number of iterations.
The same quality of solutions is achieved 5.4 times faster on the GPU.

Results for the J120 data-set are shown in Tables \ref{tab:comparisonQualityJ120} and \ref{tab:comparisonSpeedupJ120}.
It can be noted that the quality of GPU solutions is substantially lower for 10000 iterations.
The GPU requires about 50000 iterations to achieve the quality of the CPU solutions.
On the other hand, the GPU is able to compete with the CPU since 50000 iterations is performed 2.5 times faster than 10000 iterations for the parallel CPU version.
The GPU evaluates more than one million schedules per second, whereas the CPU evaluates one hundred thousand.

The quality of the solutions is compared with the existing solutions for the RCPSP in Table~\ref{tab:comparisonOther}.
Our proposed PTSG outperforms other Tabu Search implementations with respect to the quality of solutions.
For example, Artigues' Tabu Search \cite{Artigues2003249} has been given at least 11000 iterations for the J120 data-set and achieves 36.16\,\% CPM dev.
Having had 10000 iterations the proposed PTSG reaches 33.41\,\% and 34.67\,\% for the CPU and GPU respectively.
In addition, the proposed PTSG can be just as good as other heuristic approaches like Ant Colony Optimization and Simulated Annealing.
On the other hand, the state of the art random-key genetic algorithms give even better solutions than the PTSG.

From the performance point of view it is difficult to compare since the different algorithms and hardware architectures were used for experiments.
For example, Artigues's Tabu Search requires 67\,s per J120 instance on average. The testing configuration was not stated.
The PTSG requires 4.9\,s (10000 iterations) on the mid-range GPU with the substantially higher quality of solutions.
The genetic algorithm by Gon\c{c}alves et al. \cite{JoseGAFBI} takes 180\,s per J120 instance on average on the Intel Core 2 Duo 2.4 GHz processor.

\subsection{Evaluation of the Selection Heuristic}

The heuristic (see Section \ref{sec:heurEval}) is using the JRip classifier from the Weka data mining tool~\cite{Weka} to decide which resources evaluation algorithm should be faster.
To get data for the learning, the Progen generator \cite{Kolisch98benchmarkinstances} was used to generate 4 data-sets with 30, 60, 90, and 120 activities respectively.
The parameters of the generated data-sets were set the same as for J30, J60, J90, and J120 data-sets with the exception that different random seeds were used.
For each generated data-set the classifier was learned using \verb+weka.classifiers.rules.JRip -F 3 -N 2.0 +\linebreak[4]
\verb+-O 10 -S 0+ command and tested on the corresponding standard data-set with the same number of activities.
The achieved results in Table~\ref{tab:classError} reveal that the accuracy is decreasing with the number of activities.
The reason for this behavior can be the smaller ratio of the resources evaluation time to the total run-time.

\begin{table}[hb]
	\centering
	\scalebox{0.9}{
	\begin{tabular}{|c|c|c|c|} \hline
		\textbf{J30} & \textbf{J60} & \textbf{J90} & \textbf{J120} \\ \hline
		72.3\,\% & 85.8\,\% & 91.9\,\% & 96.3\,\% \\ \hline
	\end{tabular}}
	\caption{Accuracy of the Selection Heuristic --- the percentage of correctly classified.}
	\label{tab:classError}
\end{table}

To prove that the proposed heuristic also improves the PTSG performance the run-time was measured for each evaluation algorithm
and normalized with respect to the reference run-time, i.e. the run-time achieved by using the heuristic.
The results in Table~\ref{tab:heurSpeed} show that the heuristic accelerates the PTSG up to 2 times
and its effect is decreasing as the evaluation of schedules becomes a less time-consuming part of the PTSG.
The time-indexed algorithm seems to be faster than the capacity-indexed algorithm on the standard data-sets.
On the other hand, the achieved speedup is dependent on the characteristics of instances and it cannot be generally determined which evaluation algorithm is faster.
The capacity-indexed evaluation algorithm is usually faster for long schedules with low resource capacities in contrast to the time-indexed algorithm which usually performs
better for short schedules with high resource capacities.

\begin{table}[ht]
	\centering
	\scalebox{0.9}{
	\begin{tabular}{|l|*4{c|}} \cline{2-5}
		\multicolumn{1}{l|}{} & \textbf{J30} & \textbf{J60} & \textbf{J90} & \textbf{J120} \\ \hline
		time-indexed & 1.02 & 1.10 & 1.09 & 1.23 \\ \hline
		capacity-indexed & 1.27 & 1.34 & 1.96 & 1.73 \\ \hline
		heuristic & 1 & 1 & 1 & 1 \\ \hline
	\end{tabular}}
	\caption{Effect of the heuristic on the PTSG performance.}
	\label{tab:heurSpeed}
\end{table}

\subsection{Demonstration of Convergence}

\begin{figure}[hb]
	\begin{center}
		\includegraphics[trim = 6mm 11.5mm 18mm 14mm, clip, width=0.5\textwidth]{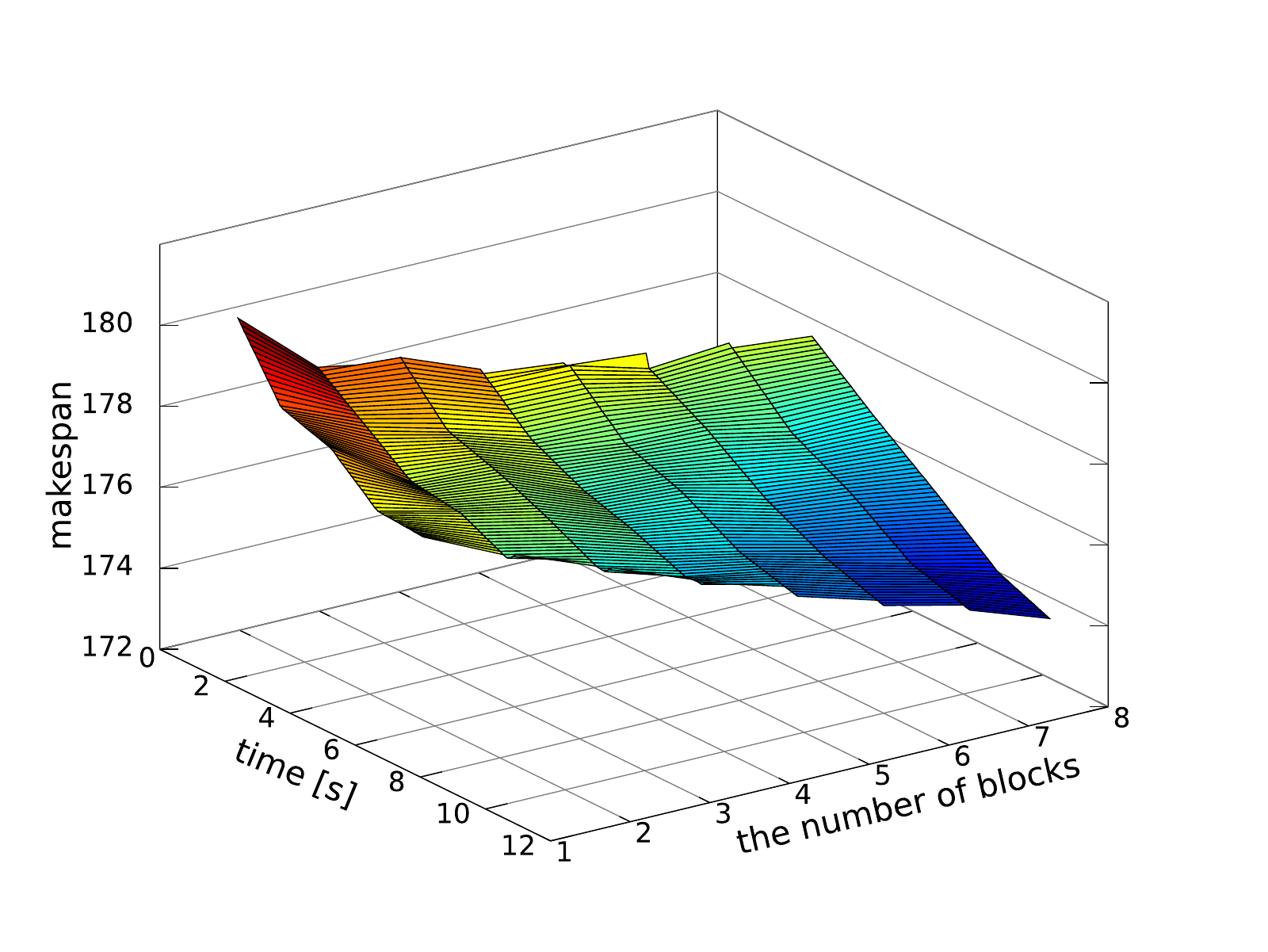}
	\end{center}
	\caption{Graph of convergence for the GPU version.}
	\label{fig:gpuConv}
\end{figure}

To demonstrate that co-operation among blocks is beneficial the graph of convergency (in Figure~\ref{fig:gpuConv}) was created for \verb+j1206_4.sm+ instance.
It can be seen that the quality of solutions is getting better with the increasing number of launched blocks, 
therefore, it is obvious that co-operation leads to the better solutions.
To ensure the smoothness of the graph each point was averaged over 50 measurements.

\section{Conclusion}

The first known GPU algorithm dealing with the Resource Constrained Project Scheduling Problem has been proposed.
The performed experiments on the standard benchmark instances reveal the merits of the proposed solution.
The achieved quality of solutions is very good and outperforms the other Tabu Search implementations to the best of our knowledge.
In addition to this, the GPU algorithm design has proved to be very effective since the mid-range GPU was substantially faster than the optimized parallel CPU version.
The Nvidia Geforce GTX 650 Ti GPU is able to evaluate more than one million schedules per second for the J120 data-set on average.
The achieved performance boost could not be reached without effective structures and auxiliary algorithms.
The Simple Tabu List implementation is adapted to the features of the GPU, the capacity-indexed evaluation algorithm was proposed and many parallel reductions were applied.
In addition to this, the homogeneous model reduces the required communication bandwidth between the CPU and the GPU.

In spite of the fact that GPUs are not primarily designed for solving combinatorial problems the rising interest about these solutions can be seen \cite{Brodtkorb20134}.
The reason for this is the high computational power of graphics cards and the relatively user friendly programming API that the CUDA offers.
So it can be expected that GPUs will be more and more used in operations research in the future.

\bibliographystyle{elsarticle-num}
\bibliography{reference}

\end{document}